\documentclass[aps,preprint,amsfonts,showpacs,superscriptaddress]{revtex4} 
\pdfoutput=1
\usepackage{bm}
\usepackage{amsmath}
\usepackage{amssymb}
\usepackage{graphicx}           
\usepackage{epsfig}
\usepackage{color}

\newcommand\nn{\nonumber}
\newcommand\bea{\begin{eqnarray}}
\newcommand\eea{\end{eqnarray}}
\newcommand\f{\frac}
\newcommand\p{\partial}
\newcommand\la{\langle}
\newcommand\ra{\rangle}

\newcommand\etal{{\emph{et al~}}}
\newcommand\g{\gamma}

\begin{document}

\title{ Heat  conduction in the $\alpha-\beta$ Fermi-Pasta-Ulam  chain}

\author{Suman G. Das}
\affiliation{Raman Research Institute, Bangalore 560080, India}

\author{Abhishek Dhar} 
\affiliation{International centre for theoretical sciences, TIFR, Bangalore 560012, India}

\author{Onuttom Narayan}
\affiliation{Department of physics, University of California, Santa Cruz, California 95064, USA}

\date{\today}

\begin{abstract}
Recent simulation results on heat conduction in a one-dimensional chain
with an asymmetric inter-particle interaction potential  and no onsite
potential found non-anomalous heat transport in accordance to Fourier's
law. This is a surprising result since it was long believed that heat
conduction in one-dimensional systems  is in general  anomalous in the
sense that the thermal conductivity diverges as the system size goes
to infinity.  In this paper we report on detailed numerical simulations
of this problem to investigate the possibility of a finite temperature
phase transition in this system. Our results indicate that the unexpected
results for asymmetric potentials is a result of insufficient chain
length, and does not represent the asymptotic behavior.
\end{abstract}

\maketitle
\newpage

\section{Introduction}
It is now  generally believed that heat conduction in one-dimensional
($1D$) momentum conserving systems is anomalous \cite{LLP03,dhar08}. 
 This  belief is supported by a large number of
 simulation studies \cite{simulations} and analytic
 work, mostly based on linear response theory
 \cite{delfini06,narayan,wang04,pereverez03,lukkarinen,BBO,beijeren}.
In the absence of an external pinning potential, as is the case  in most
realistic situations, one finds that Fourier's law is not  satisfied. One
of the predictions from Fourier's law is about the scaling form of heat
current $J$
 with system size $N$ for a system with a fixed applied temperature
 difference.  Fourier's law predicts $J \sim 1/N$ but for  one dimensional
 ($1D$) momentum conserving systems one finds:
\bea
J \sim \f{1}{N^{1-\alpha}} ~~~~~\alpha > 0~.
\eea
In the linear response regime where a small temperature difference
$\Delta T$ is applied,  one can define the conductivity through
the relation $\kappa = JN/\Delta T$. For anomalous systems one then
expects the thermal conductivity to diverge with system size as $\kappa
\sim N^\alpha$. Determining the exact value of the exponent $\alpha$
and identifying universality classes has been an outstanding problem
on which there is no consensus so far, though most numerical studies
indicate a value of $\alpha$ in the range $0.3-0.5$. In two dimensions
a logarithmic divergence of $\kappa $ with system size is expected.

Two recent papers \cite{zhao12,zhao13}  considered heat conduction in
several $1D$ models where the inter-particle interaction potential $V(x)$
is taken to be asymmetric, in the sense that $V(x) \neq V(-x)$. Based
on nonequilibrium  simulations as well as equilibrium Green-Kubo
type computations they conclude that, in certain parameter regimes,
Fourier's law {\emph{ is satisfied}} in these systems, \emph{i.e}  for
these systems $\kappa$ converges to a size-independent value. This is
a very surprising result which raises a few questions:

(a) Is there something wrong with  the analytical predictions based on
mode-coupling theory and hydrodynamic arguments?

(b) Zhong {\emph {et al}} \cite{zhao12} find normal transport at low temperatures and
anomalous transport at high temperatures. Is there a nonequilibrium
phase-transition in this system as a function of temperature, or are
finite size effects stronger at low temperatures, so that the true
asymptotic (anomalous) behavior is only seen for much larger system sizes?

In this paper we try to answer the second  question and, to some extent
the first,  by performing a detailed simulational study of heat conduction
in the so-called FPU$-\alpha-\beta$  model; although Ref.~\cite{zhao12}
states that the asymmetry in the interparticle potential is weak for this
model, we will see that this is actually an advantage. We will present
results on the size-dependence of $\kappa$ at different temperatures
and see if we can identify  a nonequilibrium transition between a
diffusive low-temperature phase and an anomalous high temperature
phase. Our results favor finite size effects as an explanation for the
low temperature behavior, and not a separate low temperature phase.
We  mostly report results based on nonequilibrium simulations but also
point out some  problems associated with the Green-Kubo approach.

One notable difference between asymmetric and symmetric potentials  is
that the former allows for thermal expansion. Zhong \etal \cite{zhao12}
observed that on applying different temperatures to the ends of a
chain of particles (with asymmetric interaction potentials), there was
a non-uniform thermal expansion in the system and this was proposed
as leading to an additional mechanism of phonon-scattering which might
somehow give Fourier-type behavior. As we discuss at the end of the paper,
this explanation is inconsistent with being in the linear response regime,
but it motivates a careful examination of thermal expansion in the system.
In the presence of thermal expansion, it becomes important to consider
boundary conditions: for a chain with fixed boundary conditions there is
no overall expansion of the whole system, while for a system with free
boundaries there is a net overall expansion. One might therefore expect
different heat transport behavior for these two boundary conditions. We
also explore this question in this paper.

\section{Results of nonequilibrium simulations} 
The FPU $\alpha-\beta$ model is described by the following  Hamiltonian:
\bea
H \! &=&\! \sum_{l=1}^{N} \f{p_l^2}{2 m} 
+ \sum_{l=1}^{N+1} 
[ k_2 \f{(q_l-q_{l-1})^2}{2} +  k_3  \f{(q_l-q_{l-1})^3}{3} + k_4 \f{(q_l-q_{l-1})^4}{4} ],~~
\eea
where $\{ q_l, p_l \}$ denote the displacement about equilibrium
lattice positions  and momenta of particles and we  have  considered
different boundary conditions (BCs).  Fixed and free BCs are obtained
by setting $q_0=q_{N+1}=0$ and $q_0=q_1, q_N=q_{N+1}$ respectively.
The interparticle harmonic spring constant is denoted by $k_2$ (which we
set to $1$) while $k_3,k_4$ denote the strengths of the cubic and quartic
interactions respectively.  To set up  heat transport in this system the
particles at the two ends of the chain are connected to stochastic white
noise heat baths at different temperatures.  The equations of motion of
the chain are then given by:
\bea
 m \ddot{q}_l =- (2 q_l-q_{l-1}-q_{l+1}) 
- k_3 [ (q_l-q_{l-1})^2 - (q_l-q_{l+1})^2 ] - k_4 [ (q_l-q_{l-1})^3 + (q_l-q_{l+1})^3 ] -\g_l \dot{q}_l 
+\eta_l~,~ \label{langevin}
\eea
with $\eta_l=\eta_L \delta_{l,1}+\eta_R \delta_{l,N},~\g_l=\g (\delta_{l,1}+
\delta_{l,N})$, 
 and where the noise terms satisfy the  fluctuation dissipation 
relations $\la \eta_L(t) \eta_L(t') \ra = 2 \g k_B T_L \delta(t-t')$,
$\la \eta_R(t) \eta_R(t') \ra = 2 \g k_B T_R \delta(t-t')$, $k_B$
being Boltzmann's constant. 

The energy current on the bond between particles $l$ and $l+1$ is given by
\bea
j_l=\f{1}{2} (\dot{q}_l+\dot{q}_{l+1})~\p H/\p q_{l}~.
\eea
We will mainly be interested in the steady state heat current which is
given by $ \la j \ra = \sum_{l=1}^{N-1} \la j_l \ra/(N-1)$, where $\la
...\ra$ denotes a steady state average. For small temperature differences
$\Delta T=T_L-T_R$,  the current will vary linearly with $\Delta T$
and we define a size-dependent conductivity as
\bea
\kappa=\f{N \la j \ra}{\Delta T}~. \label{neqkap}
\eea
In general $\kappa$ will be a function of $T=(T_L+T_R)/2$ and the
parameters $k_3,k_4$.  As noted in \cite{dhar08}, Eqs.~(\ref{langevin})
are invariant under the transformation $T_{L,R} \to s T_{L,R}$, $\{ q_l\}
\to \{ s^{1/2} q_l \}$ and $(k_3,k_4) \to (k_3/s^{1/2},k_4/s)$.  This
implies the scaling relation $J(T_L,T_R,k_3,k_4)= s J(T_L/s,T_R/s,s^{1/2}
k_3,sk_4)$.  Defining the conductivity as in Eq.~(\ref{neqkap}) we then
get $\kappa(T,k_3,k_4)=\kappa(1,T^{1/2}k_3,T k_4)$. In our study we
keep $k_3, k_4$ fixed and study the effect of changing $T$. Note that
increasing the temperature is equivalent to increasing both nonlinear
terms in the interparticle potential, but the quartic term increases
much faster, and so we expect any effect of an asymmetric potential to be
more pronounced at low temperatures. This is borne out by our simulations.

\begin{figure}
\vspace{1cm}
\includegraphics[width=3.3in]{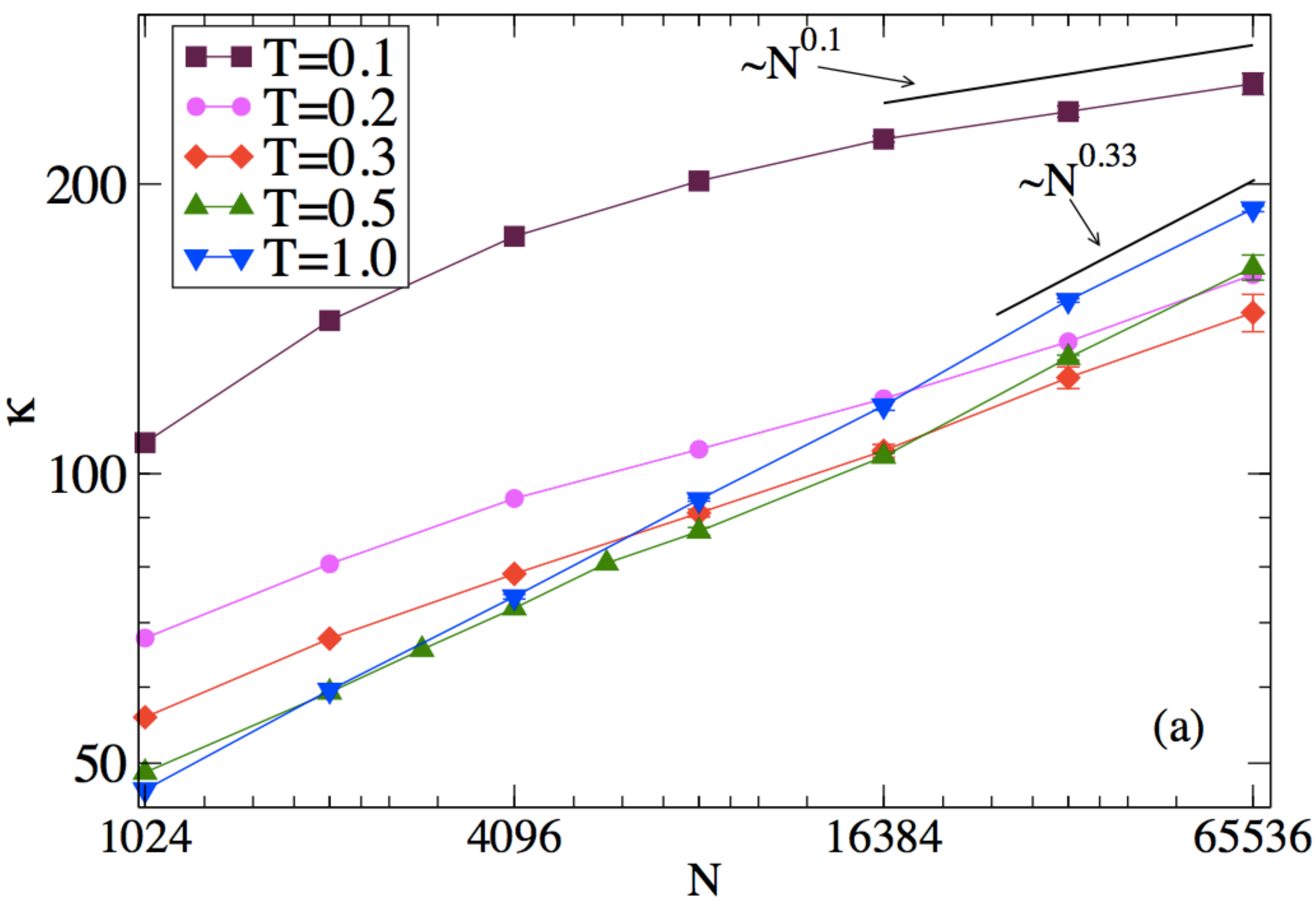}
\hspace{-0.cm}\includegraphics[width=3.3in]{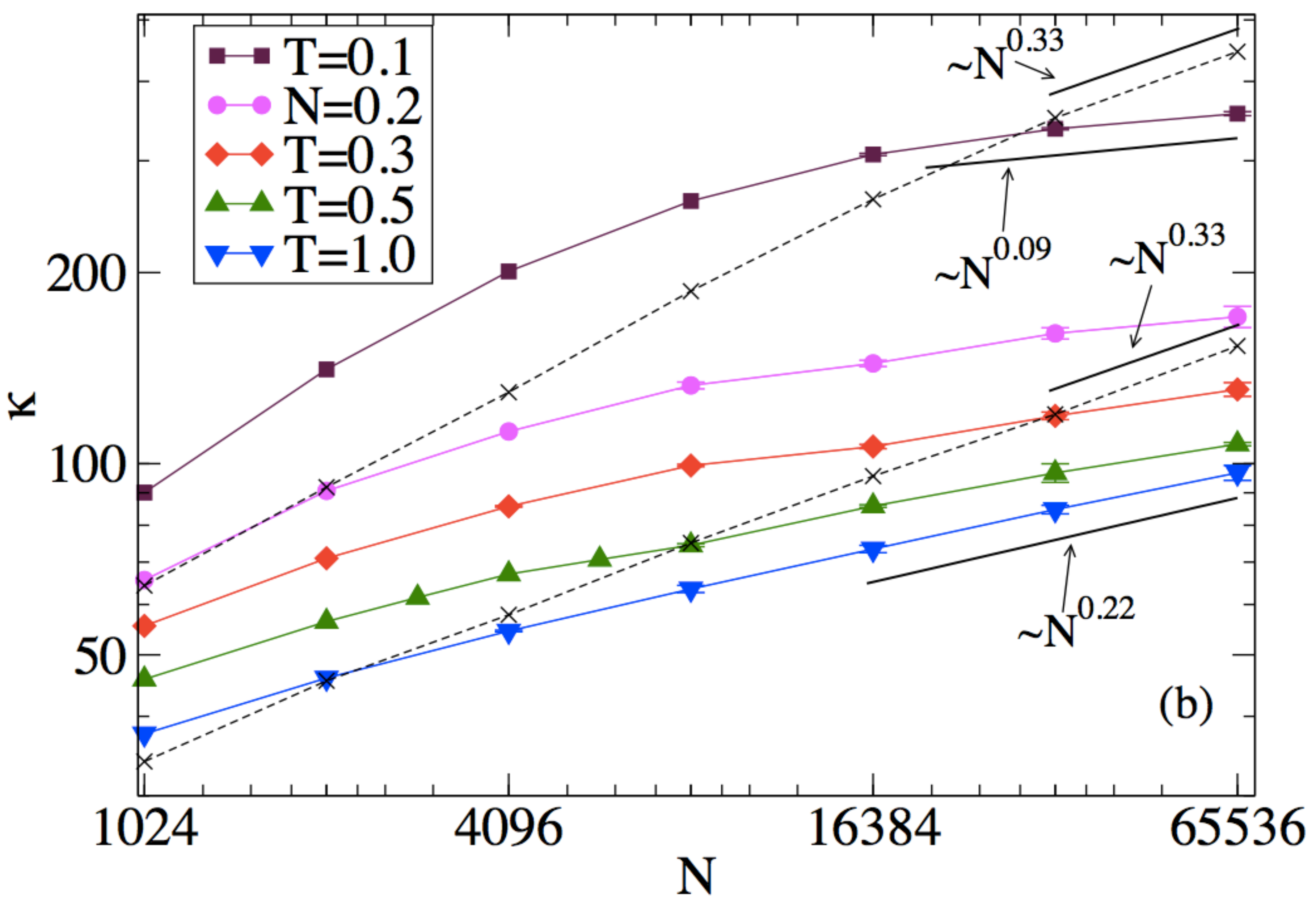}
\caption{Plot of the conductivity $\kappa$ versus
system size in the FPU~$\alpha-\beta$ model for (a)free and (b)fixed BCs at various values of the average temperature $T=(T_L+T_R)/2$ between $0.1$ and $1.0$. The parameters in the simulation were set at $k_2=k_4=1$ and $k_3=-1$. The dashed lines in (b) are results for the FPU~$\beta$ ($k_3=0$) model at temperatures $T=0.1$ (upper curve)  and $T=1.0$.} 
\label{kN}
\end{figure}

In our simulations we used the velocity-Verlet algorithm with time
steps $dt=0.005$ \cite{AT87}. We used  $O(10^9)$ steps for relaxation
and the same number of steps for  averaging.  In all our simulations
(except when otherwise mentioned) we set  $m=1,\gamma=1.0$, $k_2=k_4=1$
and $k_3=-1$. With $T_L=T+\Delta T/2$ and $T_R=T-\Delta T /2$ we obtained
data for $(T,\Delta T)=(1.0,0.5), (0.5,0.2), (0.3,0.1), (0.2,0.05),
(0.1,0.05)$. The size dependence of $\kappa$ at these temperatures is
plotted in Fig.~(\ref{kN}) for fixed and free BC. For free BC, we get the
expected $\kappa\sim N^{0.33}$ behavior at high temperatures. As has been
observed in \cite{zhao12}, the low temperature thermal conductivity seems
to saturate at large system sizes. However, we see that the $\kappa(N)$
curves at intermediate temperatures ($T=0.2, 0.3$ and 0.5) flatten out
in an intermediate range of $N$ before turning around and approaching
the $\kappa\sim N^{0.33}$ form for $T=1.0$ in the large-$N$ asymptotic
regime. As the temperature is reduced, this effect increases rapidly:
the flattening is barely noticeable for $T=0.5,$ while it is very clear
for $T=0.3$ and $T=0.2.$ For $T=0.1,$ one only sees a flattening of the
$\kappa(N)$ curve, yielding a (substantially) smaller apparent $\alpha,$
but based on our observations for $T > 0.1,$ we believe that this is
simply because the large-$N$ asymptotic regime has been pushed outside
the range of our simulations. Note that if $\kappa(N)$ were to satisfy
Fourier's law for large $N,$ we would expect the flattened regime in
the $\kappa(N)$ plots to broaden and move to the left as the temperature
is lowered but with {\it no\/} subsequent turnaround. This is not what
is seen. Although the asymmetry in the potential has greater impact at
lower temperatures, it is an irrelevant operator in the renormalization
group sense.

The results for fixed BC as the temperature is varied are qualitatively
similar but less clear: there are indications of flattening at
intermediate $N$ for $T = 0.2$ and $T=0.3,$ but because even the curve
for $T=1$ shows $\alpha < 0.33,$ the flattening is harder to discern.
For comparison we also plot the results for the FPU-$\beta$ model
($k_3=0$) where one can see the same behavior $\kappa \sim N^{0.33}$ at
both low and higher temperatures. It is thus clear that the difficulty
in convergence to the asymptotic limit is closely related to the cubic
term in the potential. It has been suggested~\cite{mendl13,spohn13}
that that the hydrodynamic equations for a chain, on which analytical
predictions are based, are different for the special case when the chain
is at zero pressure. But this corresponds to the free BC chain or the
FPU-$\beta$ model, for both of which we seem to find $\alpha = 1/3,$
whereas the {\it fixed\/} BC chain does not yield $\alpha = 0.33.$ (With
the parameters used here, the fixed BC chain has an average pressure of
$p=-0.4$ for $T=1$ and $p=-0.1$ for $T=-0.1.$) Thus although we cannot
say so conclusively, we believe that even the fixed BC chains should,
in the asymptotic large-$N$ limit, approach $\alpha = 1/3.$

\begin{figure}
\vspace{1cm}
\includegraphics[width=6.8in]{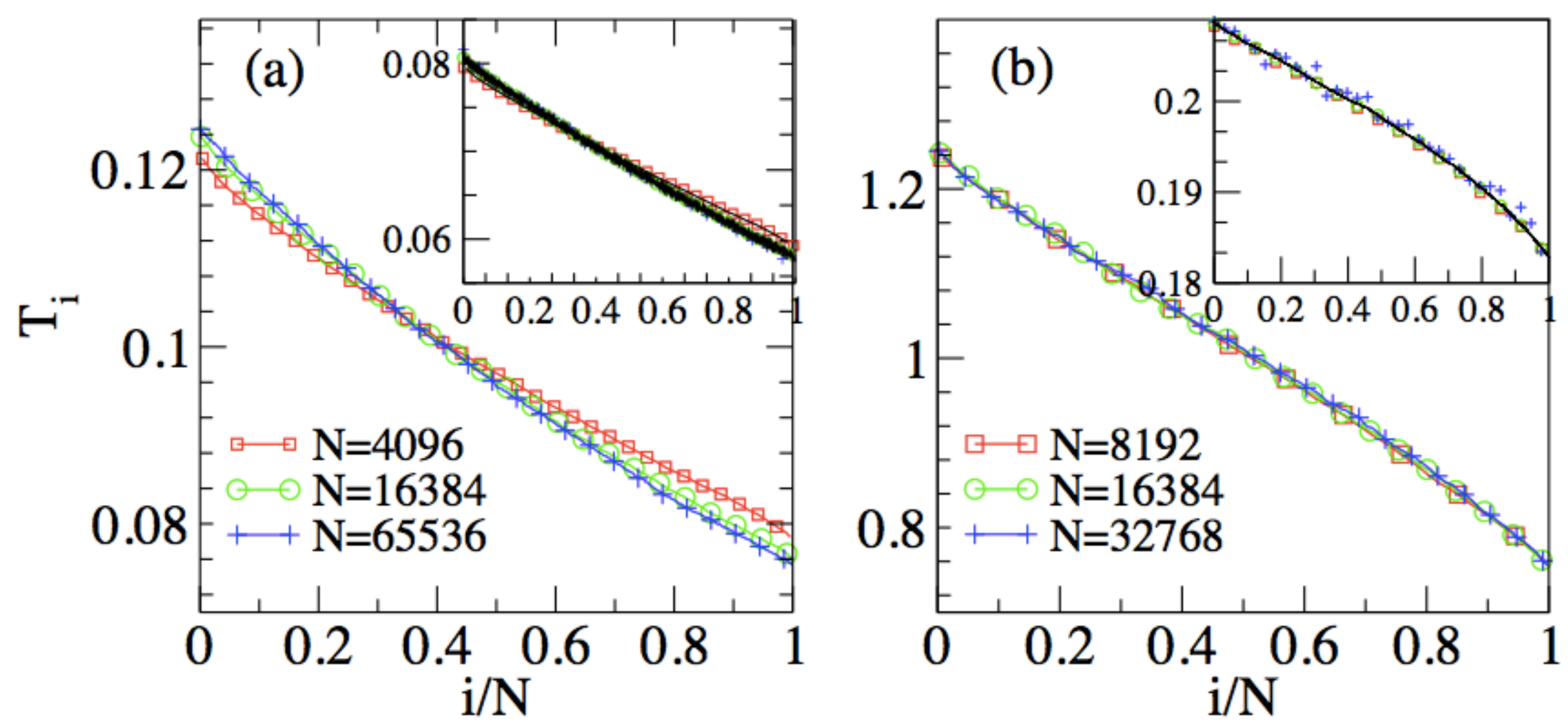}
\caption{Plot of temperature profiles for 
different system sizes for free BC at  temperatures (a) $T=0.1$ and (b) $T=1.0$. The insets show the expansion profile and the solid lines are from predictions assuming local thermal equilibrium.} 
\label{prof-free}
\end{figure}

\begin{figure}
\vspace{1cm}
\includegraphics[width=6.8in]{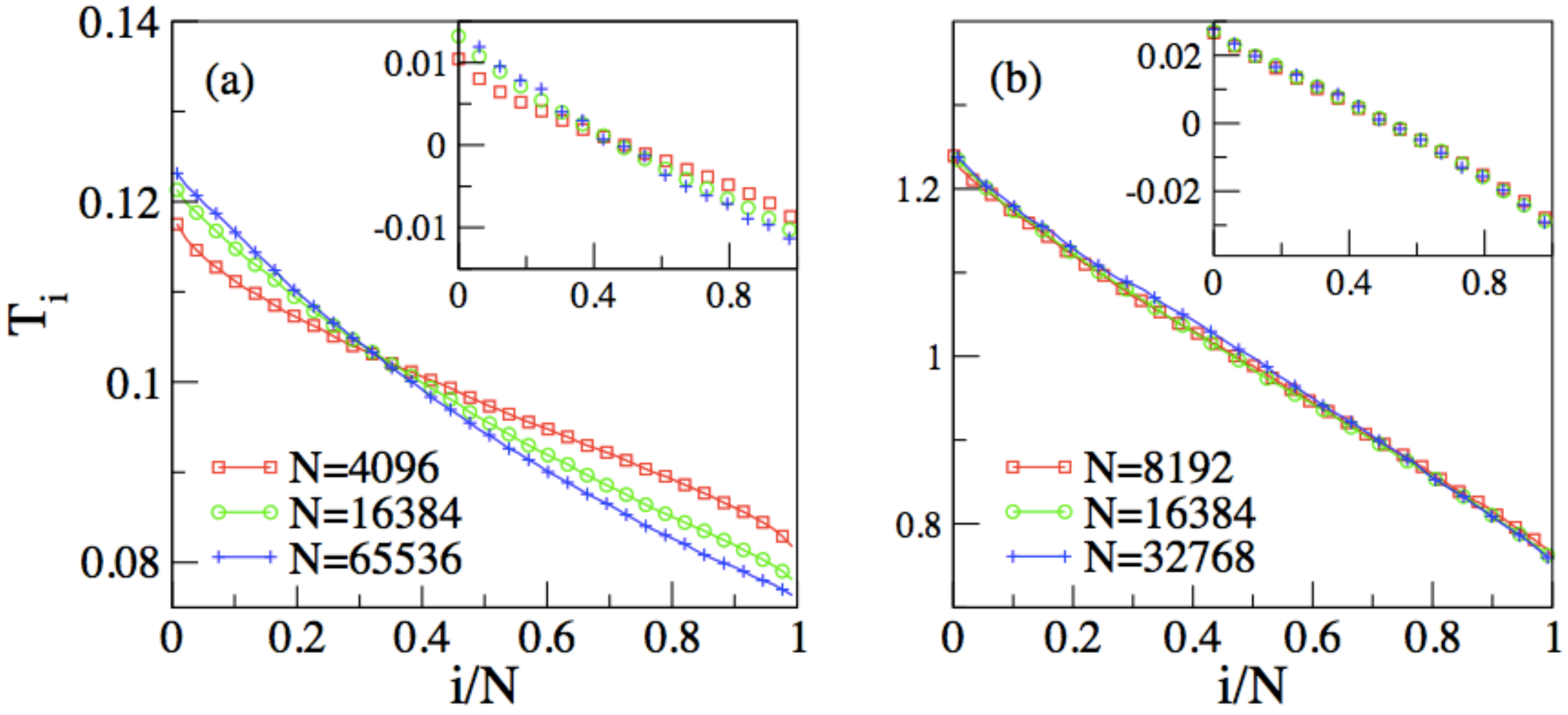}
\caption{Plot of temperature profiles for 
different system sizes for fixed BCs at  temperatures (a) $T=0.1$ and (b) $T=1.0$. The insets show the expansion profile.} 
\label{prof-fixed}
\end{figure}

We now report on other steady state properties.  The temperature
profile (defined as $T_l=m \la \dot{x}_l^2 \ra$) in the system gives
some indication as to whether one is in the hydrodynamic regime. The
temperature profiles for large $N$ should collapse and jumps  at the
boundaries should be small.  In Figs.~(\ref{prof-free},\ref{prof-fixed})
we plot the size dependence of temperature profiles for free and fixed 
BCs at low ($T=0.1$) and high ($T=1.0$) temperatures. In the insets we
also plot the profile for the thermal  expansion given by $\la r_l\ra =\la
q_{l+1}-q_l\ra$.  The most relevant observation from the temperature and
expansion profiles is that at low temperatures, even at the largest system
sizes, we do not see a collapse of the temperature profiles and this tells
us that that the asymptotic regime has not yet been reached. Secondly we
see that for free BC the convergence is faster. Both these observations
are consistent with our conclusions from the conductivity curves,
that the change seen in the thermal conductivity of the system as
the temperature is lowered does not reflect the asymptotic large-$N$
behavior in this temperature regime.

We also compare the measured local expansion to the prediction from
the measured local temperature and the assumption of local thermal
equilibrium, to see if the assumption is valid.  The  local expansion
can be obtained by assuming local thermal equilibrium and using the
local temperature.  For the case of free BC this is simple and the local
expansion $\la r_l\ra$ is given by
\bea
\la r_l \ra = \f{\int_{-\infty}^\infty dr_l ~r_l~ e^{-\beta_l V(r_l)}}{\int_{-\infty}^\infty dr_l ~e^{-\beta_l V(r_l)} }~,   
\eea
where $\beta_l=1/T_l$. For the case of fixed BC the variables are not
independent and we need  impose the constraints $q_{0}=q_{N+1}=0$ or
equivalently $q_0=0,~\sum_{i=0}^{N} r_i=0$. The local expansion is then
given by
\bea
\la r_l \ra = \f{\int_{-\infty}^\infty dr_0 \ldots \int_{-\infty}^\infty dr_{N}~ r_l~ e^{-\sum_l \beta_l V(r_l)} ~\delta (\sum_{j=0}^{N} r_j) }{\int_{-\infty}^\infty dr_0 \ldots \int_{-\infty}^\infty dr_{N} ~ e^{-\sum_l \beta_l V(r_l)}~ \delta (\sum_{j=0}^{N} r_j) }~.\nn
\eea
Introducing the Fourier representation $\delta(y)=\int_{-\infty}^\infty
dk e^{i k y}/(2 \pi)$ we then get
\bea
\la r_l \ra = \f{\f{1}{2 \pi} \int_{-\infty}^\infty dk \prod_{j\neq l} \tilde{f}_j(k) \int_{-\infty}^\infty dr_l ~r_l~ e^{i k r_l}~ e^{-\beta_l V(r_l)}}
{\f{1}{2 \pi} \int_{-\infty}^\infty dk \prod_{j} \tilde{f}_j(k)}, \label{expfix}
\eea
where $\tilde{f}_j(k) = \int_{-\infty}^\infty dr_j  e^{i k r_j}
e^{-\beta_j V(r_j)}$.  The solid black lines in the inset of
Fig.~(\ref{prof-free}) show the local equilibrium predictions for $\la
r_l \ra$.   For fixed BC it is difficult to evaluate Eq.~(\ref{expfix})
numerically for large system sizes, but for smaller sizes we find
reasonable agreement with simulation results.

\section{Results from linear response theory}
The  Green-Kubo formula relates transport coefficients of a system to
appropriate  equilibrium time correlation functions. For heat transport
in one dimension,  the Green-Kubo formula states
 \begin{equation}
 \kappa_{GK}=\lim_{\tau \to \infty} \lim_{N \to \infty} 
\frac{N}{k_B T^2 } \int_0^\tau \langle j(0)j(t) \rangle dt, \label{GK}
 \end{equation}
where  the  current operator $j= \Sigma_1^{N} j_{l,l+1}/N$ and the
angular brackets indicates an averaging over the initial equilibrium
distribution. The time evolution is Hamiltonian. The order of the
limits in Eq.(\ref{GK}) is important. $\kappa_{GK}$ should be equal to
$\lim_{\Delta T \to 0} \lim_{N \to \infty} \kappa,$ where $\kappa$ is
defined by Eq.(\ref{neqkap}).  One also expects that the precise choice
of boundary conditions in the evaluation of the correlation function
should not matter in the thermodynamic limit.

For systems whose transport coefficients diverge in the thermodynamic
limit --- as is the case here --- it is impossible to take
the $N\rightarrow \infty$ limit in Eq.(\ref{GK}): typically,
the current-current correlation decays with a power law tail for an
infinite system, thus giving a divergent conductivity. 
If instead one evaluates the integral on the right hand side of Eq.(\ref{GK}) 
for a finite system, it has been shown for a hard particle system~\cite{onuttom03} that the result
is {\it qualitatively\/} 
different with periodic BC and with open BC
connected to baths, calling into question any analysis that relies on
periodic boundary conditions; this is because the tail of the correlation 
function is cut off differently in the two cases.
For a chain, it
is usual to set the upper cut-off in the Green-Kubo integral to $\sim N/c$,
where $c$ is the speed of sound, and thereby obtain the scaling 
of the conductivity (although not its exact magnitude), but this is an ad hoc procedure.

As shown in Ref\cite{kundu09}, an exact Green-Kubo like linear response
relation  can be developed for open systems with heat bath dynamics.
It  relates the nonequilibrium $\kappa$ to the equilibrium current
autocorrelation decay in the following way:
\begin{equation}
\kappa=\lim_{\Delta T \to 0} \frac{N \langle j \rangle}{\Delta T}= \f{N}{k_B T^2} \int_0^\infty 
\langle {j(0)}{j(t)} \rangle dt, \label{GKex}
\end{equation}
where  the  current operator now is $j= \Sigma_1^{N-1}
j_{l,l+1}/(N-1)$. Formally Eq.~(\ref{GK}) and Eq.~(\ref{GKex}) have
the same structure but the correlation functions in the two cases are
computed with different dynamics. Also, Eq.~(\ref{GKex}) is well-defined
for finite systems which is not the case for Eq.~(\ref{GK}).

We note that Ref.~\cite{zhao13} and most simulations that use the
Green-Kubo formula  find the current auto-correlation  for systems
with periodic BCs. In view of the discussion above, we instead use
heat bath dynamics and fixed BCs. The heat bath dynamics corresponds
to  Eqs.~(\ref{langevin}) with $T_L=T_R=T$. We also show the results for
Hamiltonian dynamics with periodic BCs for comparison.  In our simulations
the system was initially equilibrated by connecting all points to Langevin
type heat baths at the specified temperature. After equilibration,
the heat baths were removed and the system was evolved using the two
different dynamics. In the periodic BC case, for every initial condition,
the center of mass momentum of the system was subtracted. We set the
cubic anharmonicity to a larger value ($k_3=2.0$)  in these simulations.

\begin{figure}
\vspace{1cm}
\includegraphics[width=6.8in]{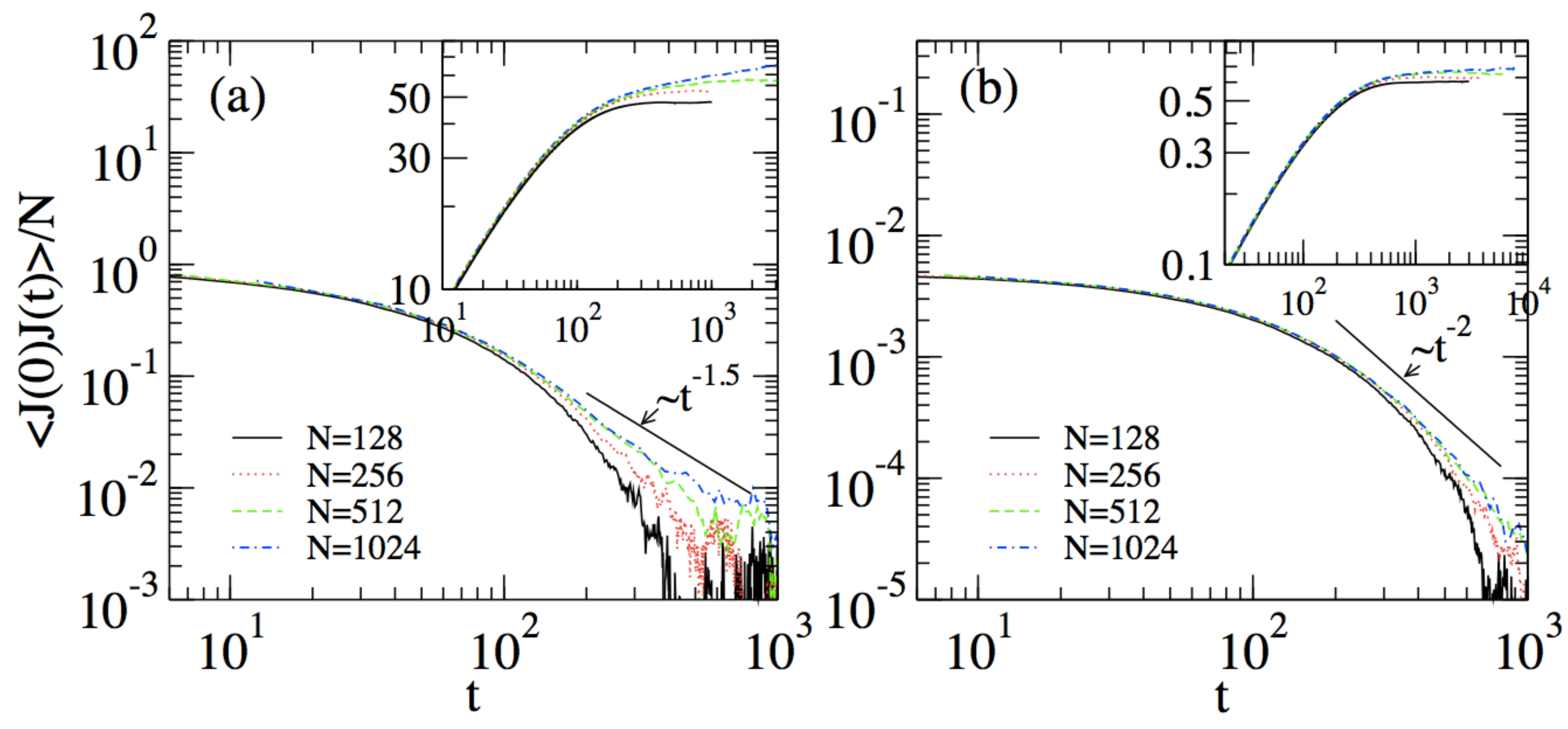}
\caption{Plot of equilibrium current-current correlations using
Hamiltonian dynamics for different system sizes  at  temperatures
(a)$T=0.1$ and (b)$T=1.0$. The insets shows the integrated correlation
function whose saturation value (divided by $T^2$) gives the thermal
conductivity as defined by Eq~(\ref{GK}). Periodic BCs were used in
this simulation.  In these simulations we set $k_3=2.0$.}
\label{GK-Hamdyn}
\end{figure}

\begin{figure}
\vspace{1cm}
\includegraphics[width=6.8in]{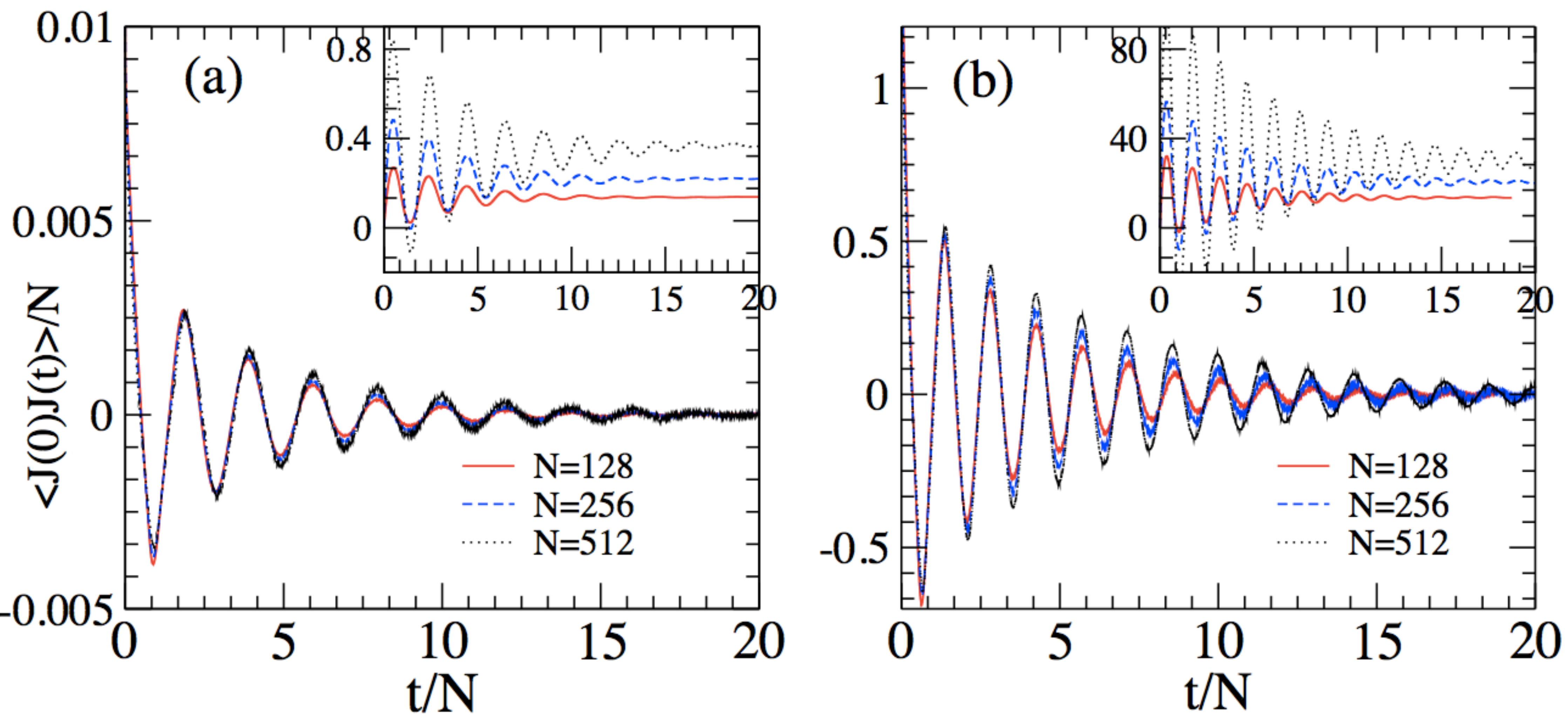}
\caption{Plot of equilibrium current-current correlations using   heat
bath dynamics for different system sizes  at  temperatures (a)$T=0.1$ and
(b)$T=1.0$. The insets show the integrated correlation function whose
saturation value (divided by $T^2$) gives the thermal conductivity as
defined by Eq~(\ref{GKex}). Fixed BCs were used in this simulation. In
these simulations we set $k_3=2.0$. }
\label{GK-HBdyn}
\end{figure}

In Figs.(\ref{GK-Hamdyn}) and (\ref{GK-HBdyn}), we show the structure of
the heat current autocorrelation function in detail for Hamiltonian and
heat bath dynamics respectively.  
Computing temporal correlation
functions from simulations is numerically challenging, and we find that
it is difficult to produce accurate data at larger system sizes. 
For the system sizes we can simulate, the form of the correlation functions
for heat bath and  Hamiltonian dynamics appear to be completely different.
In the heat bath case with fixed BC we observe
large oscillations in the correlation functions whose amplitude and
period scale with system size. These correspond to sound modes which
are reflected at the boundaries. The periodic BC correlations do not
show these oscillations and look very different. As has also been  seen in
\cite{zhao3} these, for some reason,  seem to be very small in FPU models.
Thus the difference between periodic and heat bath boundary conditions 
is even stronger than in Ref.~\cite{onuttom03}: not just the large-$t$
cutoff but the form of the heat current autocorrelation function  is different
for the two cases. Thus, although we find that the heat current autocorrelation
function decays rapidly as a function of time for periodic BC as in Ref.~\cite{zhao13},
this does not allow us to draw conclusions about the thermal conductivity.
We note in passing that the tail of the autocorrelation function is $\sim t^{-1.5}$
in the low temperature plot, which is {\it slower\/} than the $\sim t^{-2}$ at high
temperatures. 

\begin{figure}
\vspace{1cm}
\includegraphics[width=6.8in]{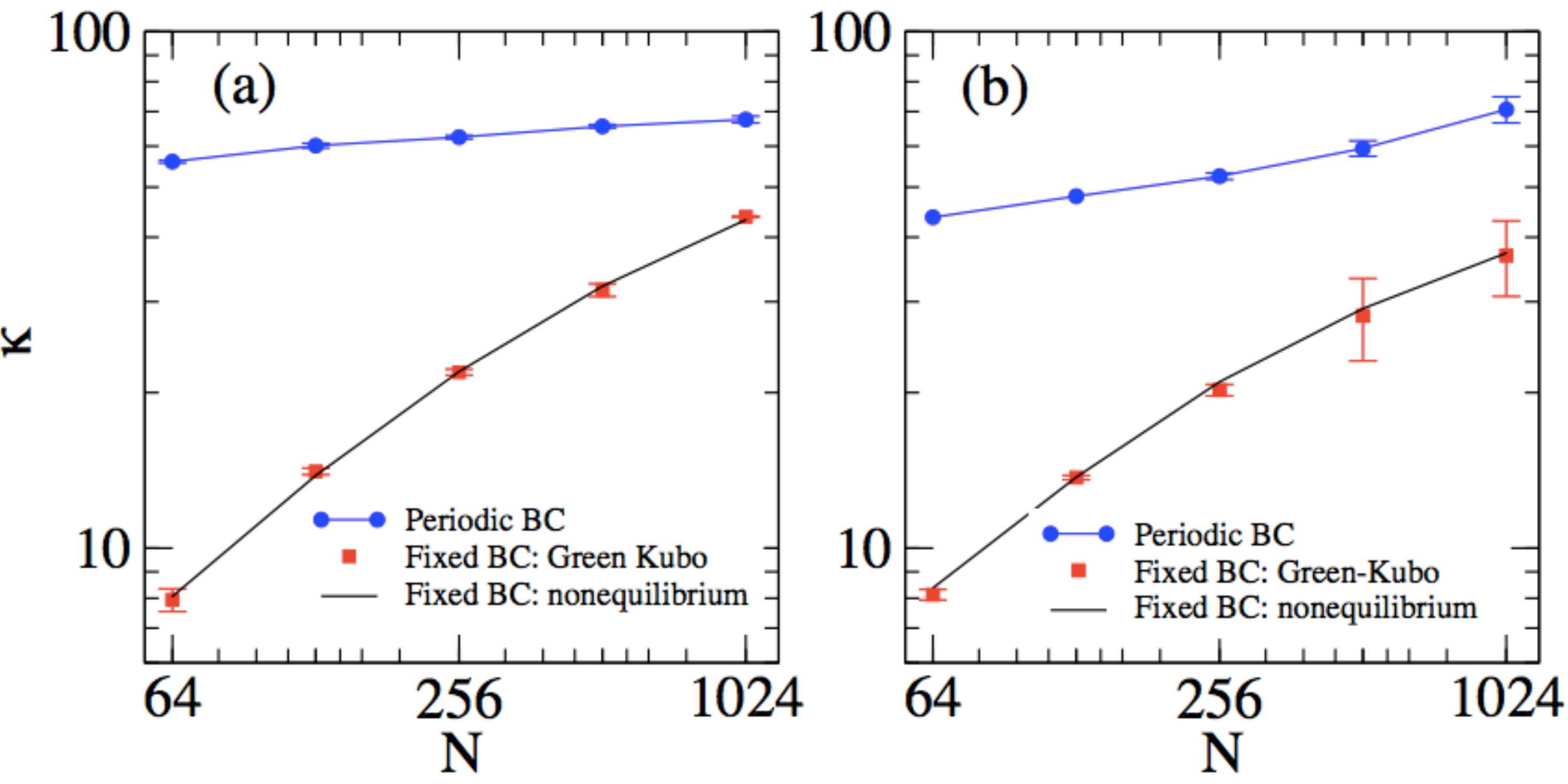}
\caption{Comparison of the thermal conductivity obtained
from the current-current correlation measurements of
Figs.~(\ref{GK-Hamdyn},\ref{GK-HBdyn}) and those obtained from direct
nonequilibrium measurements for (a)$T=0.1$ and (b)$T=1.0$~.}
\label{GK-comp}
\end{figure}

In Fig.~(\ref{GK-comp}) we plot the thermal conductivity obtained from
the heat current autocorrelation function, both for Hamiltonian and heat
bath dynamics, for different system sizes, and compare to the 
thermal conductivity directly obtained from nonequilibrium simulations. 
In the light
of the discussion above, it is not surprising that the periodic BC case,
which shows very weak $N$ dependence, is quite different from the nonequilibrium
results. The results with open BC and heat baths agree with the nonequilibrium
results, as required by Eq.(\ref{GKex}), but
nonequilibrium simulations can be performed for much larger system
sizes with relatively smaller error-bars. Therefore, we do not attempt
to obtain the asymptotic large-$N$ conductivity using the equilibrium
Green-Kubo approach.

\section{Discussion} 
In this paper, we provide numerical evidence to support the conclusion
that the asymptotic large-$N$ behavior of the thermal conductivity
for (anharmonic) chains of particles does not depend on whether the
interparticle potential is symmetric or asymmetric, contrary to recent
suggestions~\cite{zhao12,zhao13}. While the apparent thermal conductivity
at low temperatures behaves differently for the sizes we have been able
to simulate, the behavior as the temperature is tuned suggests that
this is due to insufficiently large system sizes.  However, we are not
able to rule out a phase transition in the system between diffusive
and anomalous heat transport.  We have also shown that computations
based on the Green-Kubo formula have to be performed with the proper
(i.e. heat bath) boundary conditions, in which case they agree with the
results from our nonequilibrium simulations. Computing the Green-Kubo
correlations is numerically more difficult, and one is restricted to
smaller system sizes than for the nonequilibrium simulations.

It remains an open question why, for the system sizes we are able to
simulate,  the heat conductivity at low temperatures differs so markedly
from that at higher temperatures. In Refs.~\cite{zhao12,zhao13}, it
is suggested that this may be due to inhomogeneous thermal expansion,
which gives rise to an additional scattering mechanism that reduces the
conductivity significantly (possibly due to the formation of localized
modes). While it is true that the hot and cold ends of the chain have
different thermal expansion, this cannot affect the thermal conductivity
in the linear response $\Delta T\rightarrow 0$ regime: when $\Delta T$
is small, the variation in the local thermal expansion will be $O(\Delta
T),$ and its impact on the heat current --- which would be $O(\Delta T)$
for uniform thermal expansion --- is $O(\Delta T^2).$ Equivalently, if
the inhomogeneous thermal expansion {\it does\/} affect the measured
thermal conductivity significantly, it indicates that the numerics
are not in the linear response regime. Further studies are therefore
required to determine the cause of the low temperature behavior, as well
as to shed light on the connection --- if any --- between heat current
autocorrelation functions with periodic BC and the thermal conductivity.

\section{Acknowledgements} A.D thanks DST for support through the
Swarnajayanti fellowship.

\end{document}